\documentclass[page-classic]{epl2} 

\usepackage{amsmath}
\usepackage{graphicx}
\usepackage{amssymb}

\title{The Spectrum of the Fractional Laplacian and First Passage Time Statistics}
\shorttitle{The Spectrum of the Fractional Laplacian}
\author{E. Katzav \and M. Adda-Bedia}
\shortauthor{E. Katzav \etal}

\institute{
Laboratoire de Physique Statistique de l'Ecole Normale Sup\'erieure, CNRS UMR 8550,\\
24 rue Lhomond, 75231 Paris Cedex 05, France.}


\abstract{We present exact results for the spectrum of the
fractional Laplacian in a bounded domain and apply them to First
Passage Time (FPT) Statistics of L\'evy flights. We specifically
show that the average is insufficient to describe the distribution
of FPT, although it is the only quantity available in the existing
literature. In particular, we show that the FPT distribution is not
peaked around the average, and that knowledge of the whole
distribution is necessary to describe this phenomenon. For this
purpose, we provide an efficient method to calculate higher order
cumulants and the whole distribution.}

\begin{document}

\maketitle

Anomalous diffusion is a widely investigated phenomenon with an
increasing number of applications in natural sciences
\cite{Metzler,Kantor,turbulent,Barkai,fractional,FKPZ}. Stochastic
L\'evy processes serve as a paradigm for many unusual transport
phenomena in which collective dynamics, extended heterogeneities and
other sources of jumps with a long-range distribution play an
important role and lead to anomalous diffusion. L\'evy flights are
governed by rare yet extremely large jumps of diffusing particles.
In the continuous limit, which interests us here, the L\'evy flight
process is described by a fractional diffusion equation for the
L\'evy flyer concentration field $C(x,t)$, given by
\begin{equation}
 \partial_t C( x,t) =  - \left( { - \Delta } \right)^N C( {x,t}) \, ,\qquad 0<N<1\,,
 \label{eq:C}
\end{equation}
where $\left( { - \Delta } \right)^N$ is the Riesz-Feller derivative
of fractional order $2N$, namely $\left( { - \Delta }
\right)^N=\frac{\partial ^{2N}}{\partial |x|^{2N}}$ \cite{RiFe}. Eq.~(\ref{eq:C}) describes a diffusive
process that is faster than normal diffusion (super-diffusive) \cite{Metzler}, where the
index $N$ characterizes the degree of fractality of the environment.
For $N>1$ there is no probabilistic interpretation of the equation
as a diffusion equation, although for various values of $N$
Eq.~(\ref{eq:C}) can have a different physical interpretation. For example,
$N=2$ corresponds to the overdamped vibrations of a flexible
rod, or to the out-of-equilibrium fluctuations of a slowly
growing film in molecular beam epitaxy \cite{MBE}.

Eq.~(\ref{eq:C}) has to be supplemented with appropriate Boundary
Conditions (BC) encoding the properties of the boundaries, such as
absorbing boundary conditions
\begin{equation}
 \left({-\Delta}\right)^\mu  C\left({\pm 1,t}\right) = 0 \,, \qquad 0 \le \mu  <  N/2 \, .
 \label{eq:BC}\end{equation}
Note that this condition should hold for all real values of $\mu$ in
the interval $0 \le \mu  <  N/2$. Although we will be interested
here in the case $0 < N <1$, it turns out useful to consider this
equation more generally for any $N$, and to specialize to the
desired range when drawing physical conclusions.

In the context of L\'evy flights, a classical quantity of interest
in such systems is the Mean First Passage Time (MFPT) defined as the
average time needed for a stochastically moving particle to reach
one of the two absorbing boundaries $x=\pm 1$, when it was initially
located at some point $x_0$ in the interval. It can be shown
\cite{FPT} that the First Passage Time (FPT) distribution for the
one dimensional bounded domain $\Omega$ with absorbing BC is
obtained as
\begin{equation}
 \rho \left( {t|x_0 } \right) =  - \frac{\partial }{{\partial t}}\int_\Omega {C\left( {x,t|x_0 } \right) dx}
 \label{eq:rho} \, ,
\end{equation}
where the notation includes $x_0$ giving explicit reference to the
initial condition $C\left( {x,0|x_0 } \right) = \delta \left( {x -
x_0 } \right)$. In particular, moments of the distribution $\rho
\left( {t|x_0 } \right)$ are given by
\begin{equation}
 \left\langle {t^m } \right\rangle (x_0)  = \int_0^\infty  {t^m \rho \left( {t|x_0 } \right)dt}
 \label{eq:moments} \, .
\end{equation}
The MFPT in the presence of two absorbing boundaries can be exactly
calculated using Sonin inversion formula \cite{Buldyrev}, resulting
in
\begin{equation}
 \langle t \rangle (x_0) = \frac{(1-x_0^2)^N}{\Gamma(2N+1)}
 \label{eq:MFPT} \, ,
\end{equation}
where $\Gamma(x)$ is Euler's Gamma function. This problem has been
further studied both analytically and numerically; generalized to
other boundary conditions \cite{Dybiec,Zoia}, to semi-infinite
domains \cite{Koren}, and even to complex media \cite{Condamin}.

A general question regarding the average is whether it is also
representative. A simple test is to compare the average to the
standard deviation of the distribution, or more generally to higher
order cumulants. However, so far no attention
has been given to the variance and other higher order moments of the
distribution $\rho \left(t|x_0\right)$ of FPT for L\'evy flights. Indeed,
if the distribution is peaked around the average, then
knowing the average gives a fare information about the process.
This is exactly the case for the Gaussian distribution where the
average and the variance summarizes completely the information about
the distribution. However if the distribution is not Gaussian, the knowledge of higher order cumulants
is important to characterize the process.

In the following we
show that indeed the MFPT of L\'evy flights is far from being
representative, and knowledge of higher order moments, or actually
the whole distribution, is needed to give a proper description of
the statistics of FPT. In order to achieve this task, we start by
obtaining detailed knowledge about the spectrum of the fractional
Laplacian. Using analytical and numerical results of the spectrum,
we then show that higher order moments are important in comparison
to MFPT, especially in the vicinity of the absorbing boundaries. We
also obtain the whole distribution of FPT. Finally, we comment on
how this approach can be generalized to other boundary conditions,
such as reflecting or mixed ones.

\section{Fractional Laplacian in Bounded Domains}
In a recent paper \cite{Del^N} we developed an approach to study the
spectrum of large powers of the Laplacian in bounded domains,
continuing a previous effort \cite{BHP} which focused on the ground
state (i.e., the eigenfunction corresponding to the lowest
eigenvalue). We showed that in the large $N$-limit the
eigenfunctions of $\left(-\Delta \right)^N$ with absorbing BC are
simply proportional to the associated Legendre polynomials
$P_{2N+j}^{2N}(x)$ for $j\in\mathbb{N}$, and we showed how to obtain
systematic corrections in powers of $1/N$. It turns out that this
asymptotic expansion shows remarkable convergence, so that already
for $N=1$ the asymptotic spectrum is very close to the exact one. In
addition, we showed that expressing $\left(-\Delta\right)^N$ in the
basis of the associated Legendre polynomials does not only
diagonalize it for $N\rightarrow \infty$, but is also a very useful
basis for numerically evaluating its spectrum for any $N$. However,
these results were limited to integer $N$'s and are therefore
inapplicable to the phenomena which interests us here, namely of
anomalous diffusion.

Here we show how to extend the method developed in \cite{Del^N} to
non-integer $N$'s. The challenge is to find an appropriate basis
that satisfies the BC~(\ref{eq:BC}), and coincides with the
associated Legendre polynomials for integer $N$'s. The following
normalized functions form the required basis
\begin{equation}
f_{j} \left( x \right) = {\textstyle{{\Gamma \left( {4N + 1}
\right)\sqrt {\left( {2j} \right)!\left( {2N + 2j + {\textstyle{1
\over 2}}} \right)} \left( {1 - x^2 } \right)^N } \over {4^N \Gamma
\left( {2N + 1} \right)\sqrt {\Gamma \left( {4N + 2j + 1} \right)}
}}}C_{2j}^{\left( {2N + {\textstyle{1 \over 2}}} \right)} \left( x
\right)
 \label{eq:f} \, ,
\end{equation}
for $j\in\mathbb{N}$ (including $j=0$), and $C_k^{(\lambda)}(x)$ are
the Gegenbauer polynomials \cite{abram}. In order to show this, we
need to write the matrix elements of the operator
$\left(-\Delta\right)^N$ in this basis, namely
\begin{equation}
\hat \Delta _{m,j}^N  \equiv \int_{ - 1}^1 {f_{m} \left( x
\right)\left[ {\left( { - \Delta } \right)^N f_{j} \left( x
\right)} \right]dx}  = \hat \Delta _{j,m}^N
 \label{eq:matrix} \, .
\end{equation}
This can be achieved using $\frac{d^\mu}{d|x|^\mu}x^\lambda = \frac{{\Gamma
\left( {\lambda  + 1} \right)}}{{\Gamma \left( {\lambda  - \mu  + 1}
\right)}}x^{\lambda  - \mu }$ \cite{RiFe}, and the series expansion of the Gegenbauer
polynomials \cite{abram}
\begin{equation}
C_{2j}^{(\lambda)} \left( x \right) = \sum\limits_{\ell  = 0}^j
{\frac{{\left( { - 1} \right)^\ell  (\lambda)_{2j - \ell } \left(
{2x} \right)^{2j - 2\ell } }}{{\ell !\left( {2j - 2\ell }
\right)!}}}
 \label{eq:Gegen} \, ,
\end{equation}
where $(\lambda)_n$ is the Pochhammer symbol. After some
algebra similar to \cite{Del^N}, the following expression for the
matrix elements are obtained
\begin{eqnarray}
\begin{array}{l}
 \hat \Delta _{m,j}^N  = \sqrt {{\textstyle{{\frac{\left( {2j} \right)!{\left( {4N + 2m} \right)!}}{{\left( {2m} \right)!\left( {4N + 2j} \right)!}}}}}} {\textstyle{{\sqrt {\left( {4N + 4m + 1} \right)\left( {4N + 4j + 1} \right)} \left( { - 1} \right)^{N + j} } \over {2^{2N + 1} \Gamma \left( {N + {\textstyle{1 \over 2}}} \right)}}}\sum\limits_{i = 0}^j {{\textstyle{{\left( { - 4} \right)^i \left( {2N + 2i} \right)!\Gamma \left( {2N + j + i + {\textstyle{1 \over 2}}} \right)\Gamma \left( {i + {\textstyle{1 \over 2}}} \right)} \over {\left[ {\left( {2i} \right)!} \right]^2 \left( {j - i} \right)!\Gamma \left( {N + i + {\textstyle{3 \over 2}}} \right)}}}}  \\
  \quad\quad \times {}_3F_2 \left( {\begin{array}{*{20}c}
   {i - j, - N,2N + i + j + {\textstyle{1 \over 2}}}  \\
   {i + {\textstyle{1 \over 2}},i + 1}  \\
\end{array};1} \right){}_3F_2 \left( {\begin{array}{*{20}c}
   {2N + m + {\textstyle{1 \over 2}}, - m,N + 1}  \\
   {2N + 1,N + i + {\textstyle{3 \over 2}}}  \\
\end{array};1} \right) \\
 \end{array}
 \label{eq:matrix1} \, ,
\end{eqnarray}
where ${}_3F_2 \left( {\begin{array}{*{20}c} {a,b,c}\\{d,e}\\
\end{array};x} \right)$ is the generalized hypergeometric function
\cite{abram}. Interestingly, for integer $N$'s this expression
coincides with the one obtained in \cite{Del^N}. This implies
immediately that all the results obtained for large $N$'s apply also for the fractional case, and we can readily
use the approach developed in
\cite{Del^N} with the important difference that
for non-integer $N$'s one should use as basis the
functions~(\ref{eq:f}) and not the associated Legendre polynomials.
We therefore obtain for the eigenfunctions

\begin{eqnarray}
v_j(x) &=& f_{j}(x) + \frac{\left[ {(2j)(2j-1)} \right]^{3/2}}{32N^2} f_{j - 1}(x)  \nonumber\\
 &-& \frac{\left[ {(2j + 2)(2j + 1)} \right]^{3/2}}{32N^2} f_{j +1}(x)
 + O\left( {\frac{1}{{N^3 }}} \right)
 \label{eq:eigenvector-corr} \, ,
\end{eqnarray}
and for the eigenvalues
\begin{eqnarray}
\lambda _j  &=& \sqrt 2 \frac{(4N)^{2j}}{(2j)!} \Gamma(2N+1) \left[ {1 - \frac{{3 + 4j + 8j^2 }}{{16N}} + } \right. \nonumber \\
&+& \left. {\frac{{75 + 344j + 672j^2  + 832j^3  + 192j^4 }}{{1536N^2 }} + O\left( {\frac{1}{N^3}} \right)} \right]
 \label{eq:lambdaN} \, .
\end{eqnarray}

\begin{figure}[ht]
\centerline{\includegraphics[width=7cm]{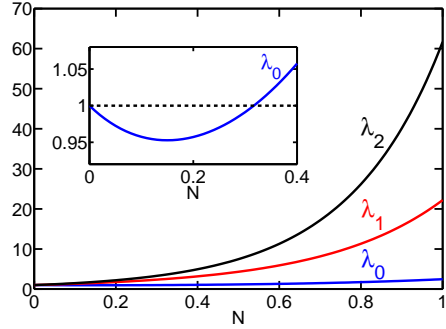}}
 \caption{The first three eigenvalues $\lambda_0(N),\lambda_1(N)$ and $\lambda_2(N)$ of $(-\Delta)^N$ in the range $0\leq N\leq1$. Note that all the eigenvalues are degenerate at $N=0$ where their value is $1$, and that only $\lambda_0$ becomes smaller than one in a certain range of $N$ (see inset).}
 \label{fig:lambdas}
\end{figure}

In particular, the numerical scheme of \cite{Del^N} will allow us to
calculate to arbitrary precision all the quantities of interest, and
thus provides a reference to the analytical results we present
below. Since we are interested in anomalous diffusion and thus
primarily on small values of $N$, we will focus from now on only on
the case $0<N\leq 1$. To demonstrate that the numerical scheme of
\cite{Del^N} can be extended to noninteger as well as small $N$'s,
we present in Fig.~\ref{fig:lambdas} the first three eigenvalues
$\lambda_0(N),\lambda_1(N),\lambda_2(N)$ in the range $0\le N \le
1$. As can be seen, the lowest eigenvalue $\lambda_0(N)$ becomes
smaller than one for values of $N \lesssim 0.32$.

In the same line of~\cite{Del^N} where a $1/N$ expansion has been
performed, we begin by studying the small $N$ behaviour of the
operator $(-\Delta)^N$. To zeroth order in $N$, it is easy to show
that Eq.~(\ref{eq:matrix1}) yields $\hat \Delta_{m,j}^N =
\delta_{mj} + O(N)$, which is consistent with the simple fact that
$\left({-\Delta} \right)^0 = 1$. Going beyond zeroth order is
cumbersome, and is mainly due to the expansion of the Generalized
Hypergeometric functions. However, we have previously shown that
(see appendix B in Ref.\cite{Del^N})
\begin{eqnarray}
&&{}_3F_2  \left( {\begin{array}{*{20}c}
   {2N + m + {\textstyle{1 \over 2}}, - m,N + 1}  \\
   {2N + 1,N + i + {\textstyle{3 \over 2}}}  \\
\end{array};1} \right) = \sum\limits_{\ell  = 0}^m {{\textstyle{{j!} \over {\ell !\left(
{m - \ell } \right)!}}}{\textstyle{{B\left( {2N + m + \ell  +
{\textstyle{1 \over 2}},{\textstyle{1 \over 2}} - \ell } \right)}
\over {B\left( {2N + m + {\textstyle{1 \over 2}},{\textstyle{1 \over
2}} - m} \right)}}}{\textstyle{{B\left( {N + 1,i + \ell  +
{\textstyle{1 \over 2}}} \right)} \over {B\left( {N + 1,i +
{\textstyle{1 \over 2}}} \right)}}}} \, ,\\
&&{}_3F_2 \left( {\begin{array}{*{20}c}
   {i - j, - N,2N + i + j + {\textstyle{1 \over 2}}}  \\
   {i + {\textstyle{1 \over 2}},i + 1}  \\
\end{array};1} \right) = {\textstyle{{4^{-2N - j - i} \sqrt \pi  \left( {2i} \right)!\left(
{j - i} \right)!} \over {\Gamma \left( {2N + j + i + {\textstyle{1
\over 2}}} \right)}}} \nonumber\\
&&\qquad \qquad \qquad \times \sum\limits_{m = 0}^{j - i}
{{\textstyle{{\left( {4N + 2j + 2m + 2i} \right)!} \over {\left( {2m
+ 2i} \right)!\left( {j - m - i} \right)!\left( {2N + j + m + i}
\right)!}}}{\textstyle{{N!} \over {m!\left( {N - m} \right)!}}}} \,
,
\end{eqnarray}
where $B(x,y)$ is the Beta function. These representations render
the small $N$ expansion straightforward yet tedious. Then, the
expansion of $\hat \Delta_{m,j}^N $ up  to first order in $N$ is
given by
\begin{equation}
\hat \Delta_{m,j}^N =
\delta_{mj} + N B_{m,j}+ O\left(N^2\right)\,,
\end{equation}
with
\begin{equation}
B_{m,j} = \left\{ \begin{array}{ll}
 D_j \,,& m = j \\
 {\textstyle{{\sqrt {\left( {4m + 1} \right)\left( {4j + 1} \right)} } \over {\left| {m\left( {2m + 1} \right) - j\left( {2j + 1} \right)} \right|}}}  \,,& m \ne j \\
 \end{array} \right.
 \label{eq:Bmj} \, ,
\end{equation}
and
\begin{eqnarray}
 D_j &\equiv& 4\psi _0 \left( {4j} \right) - 4\psi _0 \left( {2j} \right) - {\textstyle{1 \over {j\left( {4j+1} \right)}}} - 2\gamma  \nonumber \\
 &+& {\textstyle{{4j + 1} \over {2\pi }}}\sum\limits_{i = 0}^j {{\textstyle{{\left( {-4} \right)^i \Gamma \left( {j + i + {\textstyle{1 \over 2}}} \right)} \over {\left( {2i} \right)!\left( {j - i} \right)!}}}}   \sum\limits_{\ell  = 0}^j {{\textstyle{{\left( {-4} \right)^\ell
\Gamma \left( {\ell + j + {\textstyle{1 \over 2}}} \right)} \over
{\left( {2\ell } \right)!\left( {j - \ell }
\right)!}}}{\textstyle{{\left[ {\gamma  - \psi _0 \left( {i  + \ell
+ {\textstyle{3 \over 2}}} \right) + 2\psi _0 \left( {j + \ell+
{\textstyle{1 \over 2}}  } \right)} \right]} \over {\left( {\ell  +
i + {\textstyle{1 \over 2}}}
 \right)}}}}
 \label{eq:Dj} \, ,
\end{eqnarray}
where $\psi_0 (x)$ is the digamma function and $\gamma$ is the
Euler-Mascheroni constant \cite{abram}. Note that $D_0$ is obtained
by taking the limit $j \rightarrow 0$ in Eq.~(\ref{eq:Dj}), namely
\begin{equation}
 D_0 = 2(1-\gamma-\ln 2) \simeq -0.541
 \label{eq:D0} \, .
\end{equation}

Equations (\ref{eq:Bmj}-\ref{eq:D0}) allow determining the eigenvalues and the eigenfunctions of the fractional Laplacian for small $N$. Note
that the zeroth order term in the expansion of $(-\Delta)^N$ is just
the identity, implying that at the lowest order, the eigenvalues are
completely degenerate. This means that diagonalizing $(-\Delta)^N$
to first order is equivalent to the diagonalization of the matrix $B$ given by Eq.~(\ref{eq:Bmj}).
Once the eigenvalues $\{b_j\}$ and
eigenvectors $\{\overrightarrow{\beta_{j}}\}$ of the matrix $B$ are computed numerically, the eigenvalues $\{\lambda^B_j\}$ and eigenfunctions $\{v^{B}_j(x)\}$ of $(-\Delta)^N$ for small $N$ follows immediately and are just given by
\begin{eqnarray}
\lambda^B_j&=&1+N b_j\,,\\
v^{B}_j(x)&=&\sum_{n=0}^{\infty}\beta^{(n)}_j f_{n}(x)\,,
\end{eqnarray}
where $\beta^{(n)}_j$ is the $n^{th}$ component of $\overrightarrow{\beta_{j}}$. Notice that at this order,  the $N$-dependence of $v^{B}_j(x)$ comes in
only through the basis functions $f_{n}(x)$.

\begin{figure}[ht]
\centerline{\includegraphics[width=7cm]{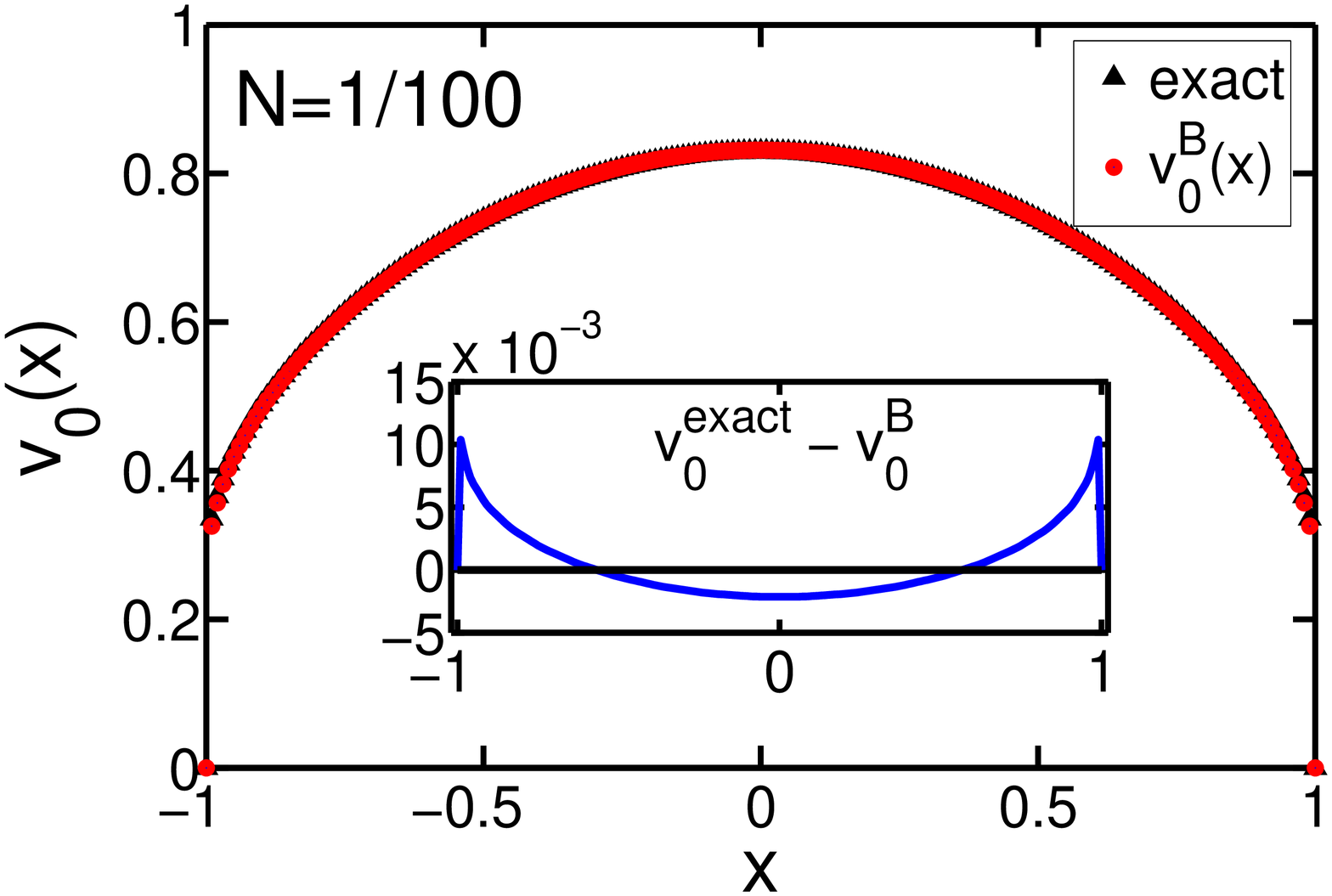}\includegraphics[width=7cm]{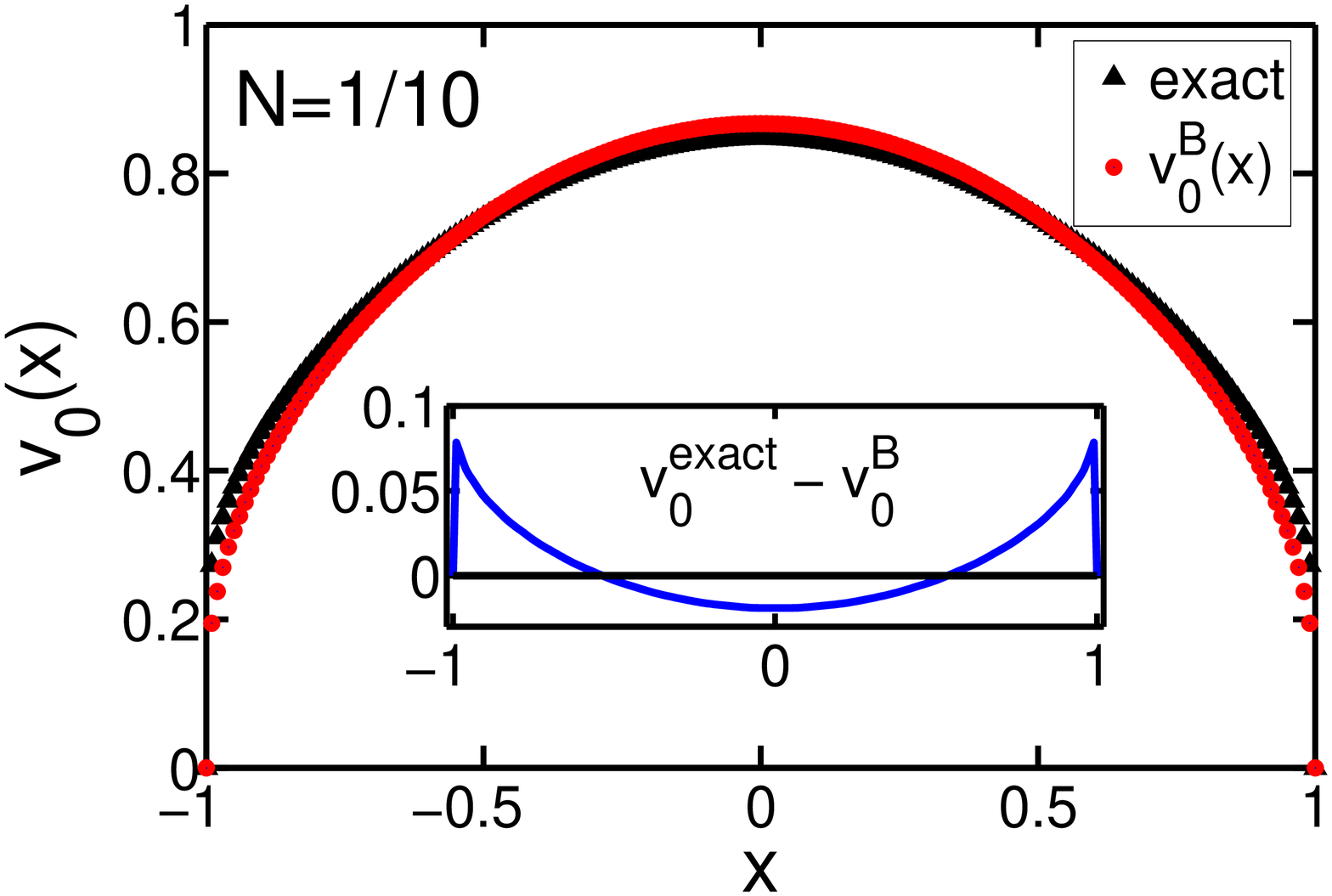}}
 \caption{A comparison of the exact ground state $v_0^{exact}(x)$ with $v_0^{B}(x)$ for various low values of $N$. The Insets show the difference $v_0^{exact}(x)-v_0^{B}(x)$.}
 \label{fig:smallN}
\end{figure}

In Table~\ref{tab:bj}, we provide the first eigenvalues of
$B_{m,j}$. Interestingly, only the smallest eigenvalue $b_0$ is
negative, implying that $\lambda^B_0<1$, consistent with the
numerical results presented in Fig.~\ref{fig:lambdas}. The other
eigenvalues $b_{j \ge 1}$ are all positive. This is in contradiction
with the small $N$ expression developed in~\cite{Zoia} for the
lowest eigenvalue, predicting $\lambda_0=1+2N(1-\gamma)>1$ (Eq.~(36)
there). The reason for this difference is twofold. First,
in~\cite{Zoia} the operator $(-\Delta)^N$ has been written using a
trigonometric basis yielding a different first-order matrix
$B_{m,j}$. For example, in the present notation, the value of $D_0$
obtained by~\cite{Zoia} is $D_0^{ZRK}=2(1-\gamma) \simeq 0.846$.
Second, the matrix $B_{m,j}$ has not been fully diagonalized
in~\cite{Zoia}, but rather approximated by $\lambda_0=1+N
D_0^{ZRK}$, predicting a positive correction for $\lambda_0$ for
small $N$'s. This suggests that the trigonometric basis used in
\cite{Zoia} is not adequate for a perturbative analysis.
Effectively, if we use the same approximation $\lambda_0=1+N D_0$ we
would obtain a negative correction, implying that the basis
$\{f_j(x)\}$ is more adapted to the operator $(-\Delta)^N$. Note
that the fact that the lowest eigenvalue is smaller than one can
have important consequences regarding the stability of the process
described by the fractional Laplacian. To allow for a more advanced
comparison, we also show  in Fig.~\ref{fig:smallN} the ground state
$v_0^B(x)$ obtained using this approximation, compared to the exact
ground state $v_0(x)$ computed numerically from the diagonalization
of~(\ref{eq:matrix1}), for various small values of $N$.

\begin{table}
\caption{The first eigenvalues of the matrix $B$ given by Eq.~(\ref{eq:Bmj}) - denoted
$b_j$. Note that the lowest eigenvalue is negative ($b_0<0$), while
all the rest are positive ($b_{j \ge 1}>0$).} \label{tab:bj}
\begin{center}
\begin{tabular}{{c}|@{}*{7}{c}}
 $j$   & $0$      & $1$     & $2$     & $3$     & $4$     & $5$ \cr
\hline
 $b_j$ & $-0.688$ & $2.727$ & $3.909$ & $4.646$ & $5.183$ & $5.606$
\end{tabular}
\end{center}
\end{table}

\begin{figure}[ht]
\centerline{\includegraphics[width=7cm]{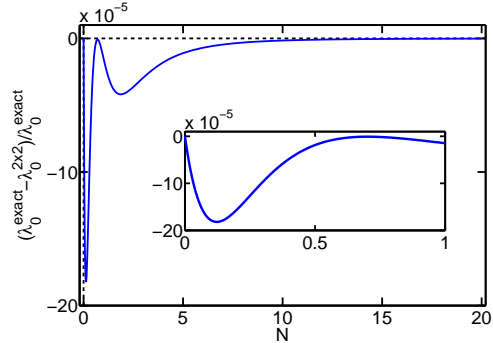}} \caption{A
comparison of the smallest eigenvalue $\lambda_0^{exact}$ with
$\lambda _0^{2\times 2}$ resulting from the $2 \times 2$
approximation  for various values of $N$. Note that the relative
error is always smaller than $2 \cdot 10^{ - 4}$ (see Inset) and
that $\lambda _0^{2 \times 2}$ becomes exact when $N \to \infty $
\cite{Del^N}.}
 \label{fig:anyNlam0}
\end{figure}

\begin{figure}[ht]
\centerline{\includegraphics[width=7cm]{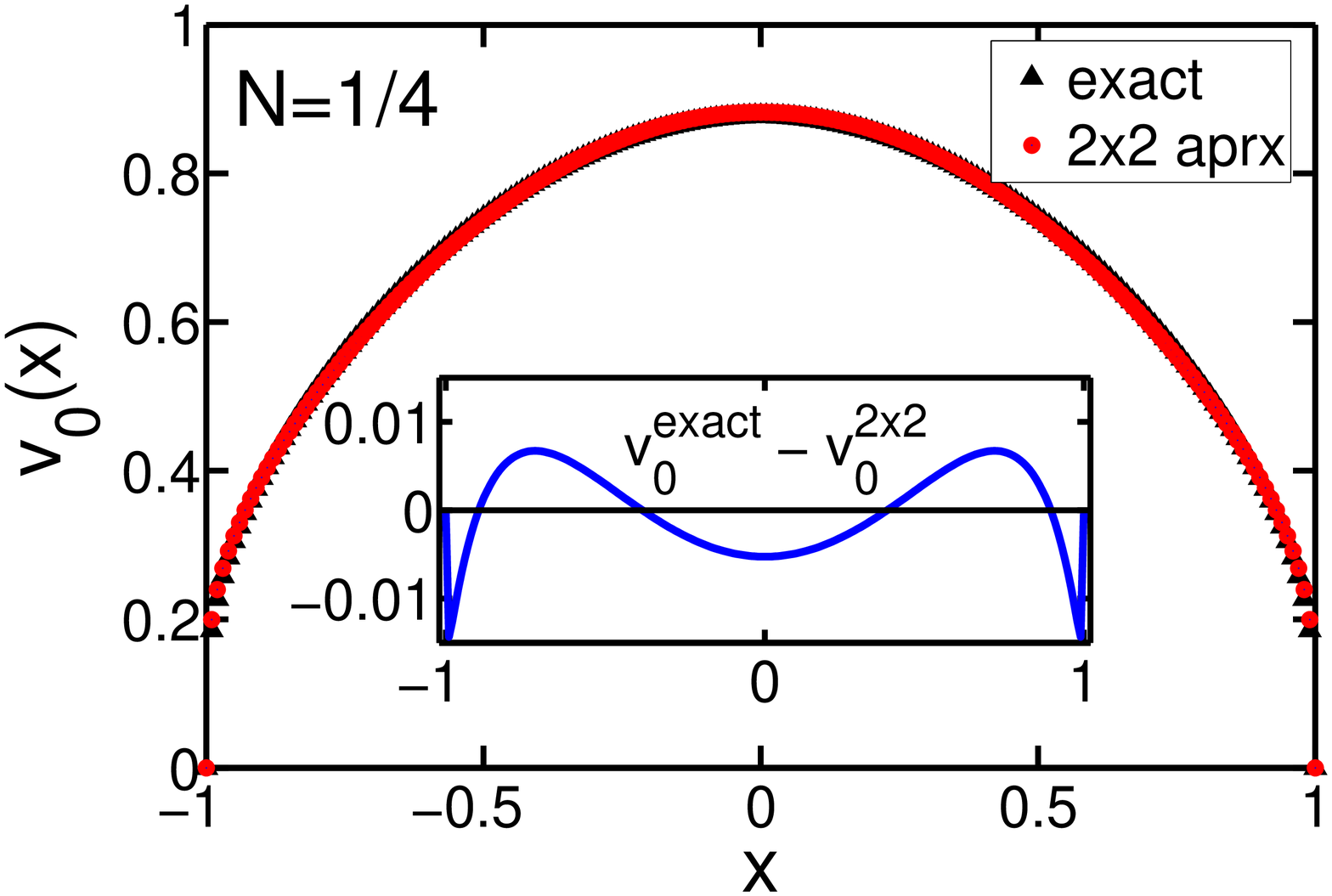}\includegraphics[width=7cm]{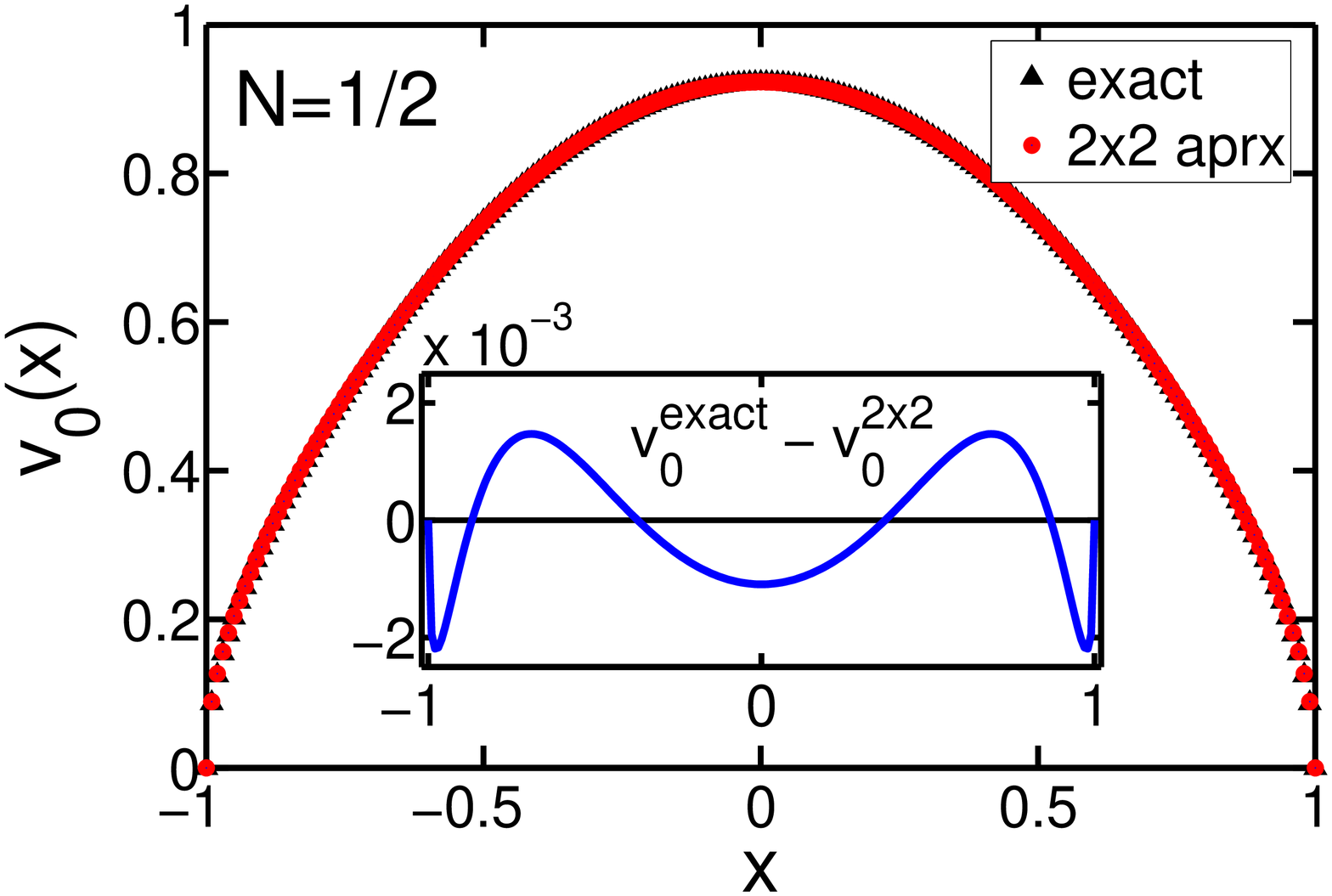}}
\centerline{\includegraphics[width=7cm]{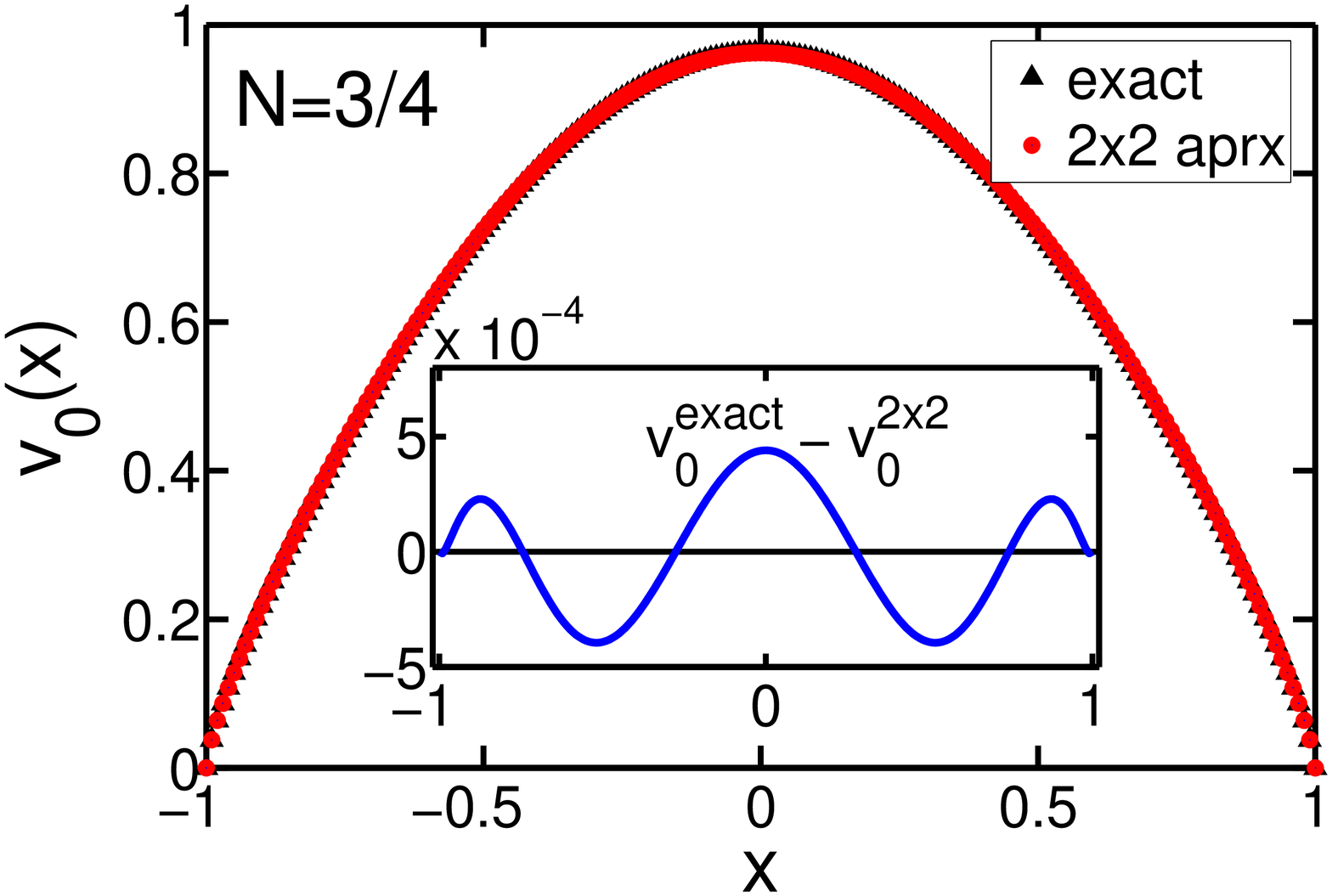}} \caption{A
comparison of the exact ground state $v_0^{exact}(x)$ with $v_0^{2
\times 2}(x)$ resulting from the $2 \times 2$ approximation  for
various values of $N$. The Insets show the difference
$v_0^{exact}(x)-v_0^{2 \times 2}(x)$.}
 \label{fig:anyNv0}
\end{figure}

Complementary to the large-$N$ study of Ref.~\cite{Del^N} and the
small-$N$ study presented above, we now develop a useful approach
for obtaining the ground state directly from the matrix $\hat
\Delta_{i,j}^N$, without expanding around a particular value of $N$.
Often in physical problems the ground state is the most interesting
quantity, as will be shown to be the case for FPT too. For this
purpose, we approximate the matrix $\hat \Delta _{m,j}^N$ as given
by Eq.~(\ref{eq:matrix1}) so that the lowest eigenvalue is given by
diagonalizing its upper $2\times 2$ sub-matrix. We then find for the
ground state
\begin{eqnarray}
\lambda _0^{2 \times 2} \left( N \right)& = &\frac{{\Gamma \left( N
\right)\Gamma \left( {N + 1} \right)\Gamma \left( {4N + 2}
\right)}}{{2^{2N + 1} \Gamma \left( {2N} \right)\Gamma \left( {2N + 2} \right)}}ab
 \label{eq:anyNlam0} \, ,\\
 v_0^{2 \times 2} \left( x \right) &=& \left[ {f_0 \left( x \right) +
af_1 \left( x \right)} \right]/\sqrt {1 + a^2 }
 \label{eq:anyNv0} \, ,
\end{eqnarray}
where
\begin{eqnarray}
 a &=& \left[ (c+1) - \sqrt {(c-1)^2 + 4b^2} \right]/2b \, , \nonumber \\
 b &=& \frac{N}{2N + 3} \sqrt{\frac{4N + 5}{2N + 1}} \, , \label{eq:abc} \\
 c &=& \frac{(4N+5)(16N^4  + 40N^3  + 42N^2  + 18N + 3)}{(2N + 1)(2N + 3)(2N + 5)}
 \, .\nonumber
\end{eqnarray}
As can be seen in Figs.~\ref{fig:anyNlam0}-\ref{fig:anyNv0}, these
expressions compare well with the exact numerical diagonalization of
$\hat \Delta_{m,j}^N$.

\section{First Passage Times}
Equipped with the new results concerning the spectrum of the
fractional Laplacian, we can readdress the statistics of FPT of
L\'evy flights. It is easy to solve formally Eq.~(\ref{eq:C}) as
\begin{equation}
 C\left( {x,t|x_0 } \right) = \sum\limits_{j = 0}^\infty  {e^{ - \lambda _j \left( N \right)t} v_j \left( x \right)v_j \left( {x_0 } \right)}
 \label{eq:solC} \, ,
\end{equation}
where here too $\left\{v_j(x) \right\}$ are the eigenfunctions of
$(-\Delta)^N$. Using Eq.~(\ref{eq:rho}) we obtain
\begin{equation}
 \rho \left( {t|x_0 } \right) = \sum\limits_{j = 0}^\infty  {\left({\int_{ - 1}^1 {v_j \left( x \right)dx} } \right)\lambda _j \left( N \right)v_j \left( {x_0 } \right)e^{ - \lambda _j \left( N \right)t} }
 \label{eq:solrho} \, ,
\end{equation}
and Eq.~(\ref{eq:moments}) for the moments becomes
\begin{equation}
 \left\langle {t^m } \right\rangle (x_0) =
\Gamma \left( {m + 1} \right)\sum\limits_{j = 0}^\infty
{\frac{{\left( {\int_{ - 1}^1 {v_j \left( x \right)dx} }
\right)}}{{\left[ {\lambda _j \left( N \right)} \right]^m }}v_j
\left( {x_0 } \right)}
 \label{eq:solmom} \, .
\end{equation}

\begin{figure}[ht]
\centerline{\includegraphics[width=7cm]{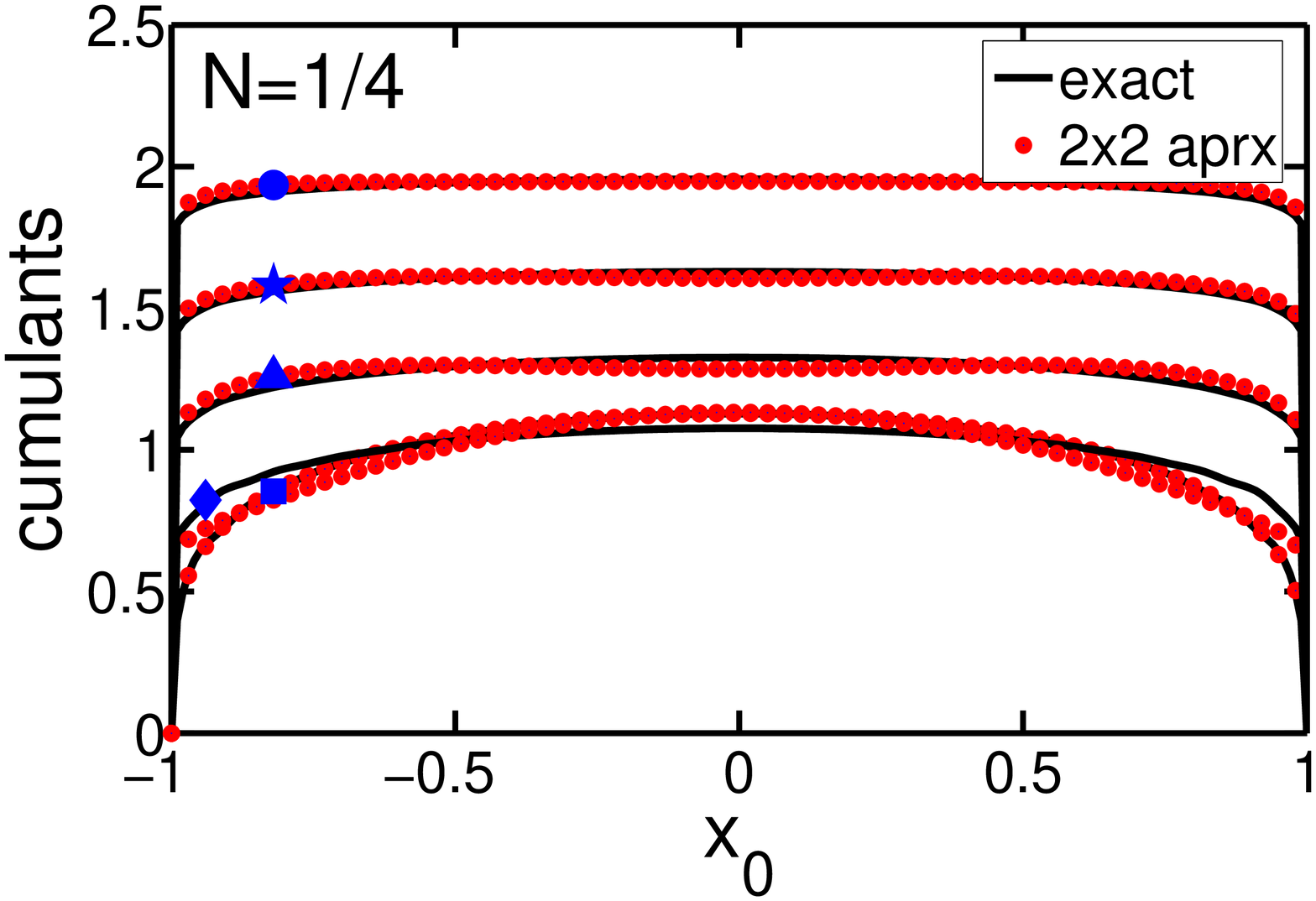}\includegraphics[width=7cm]{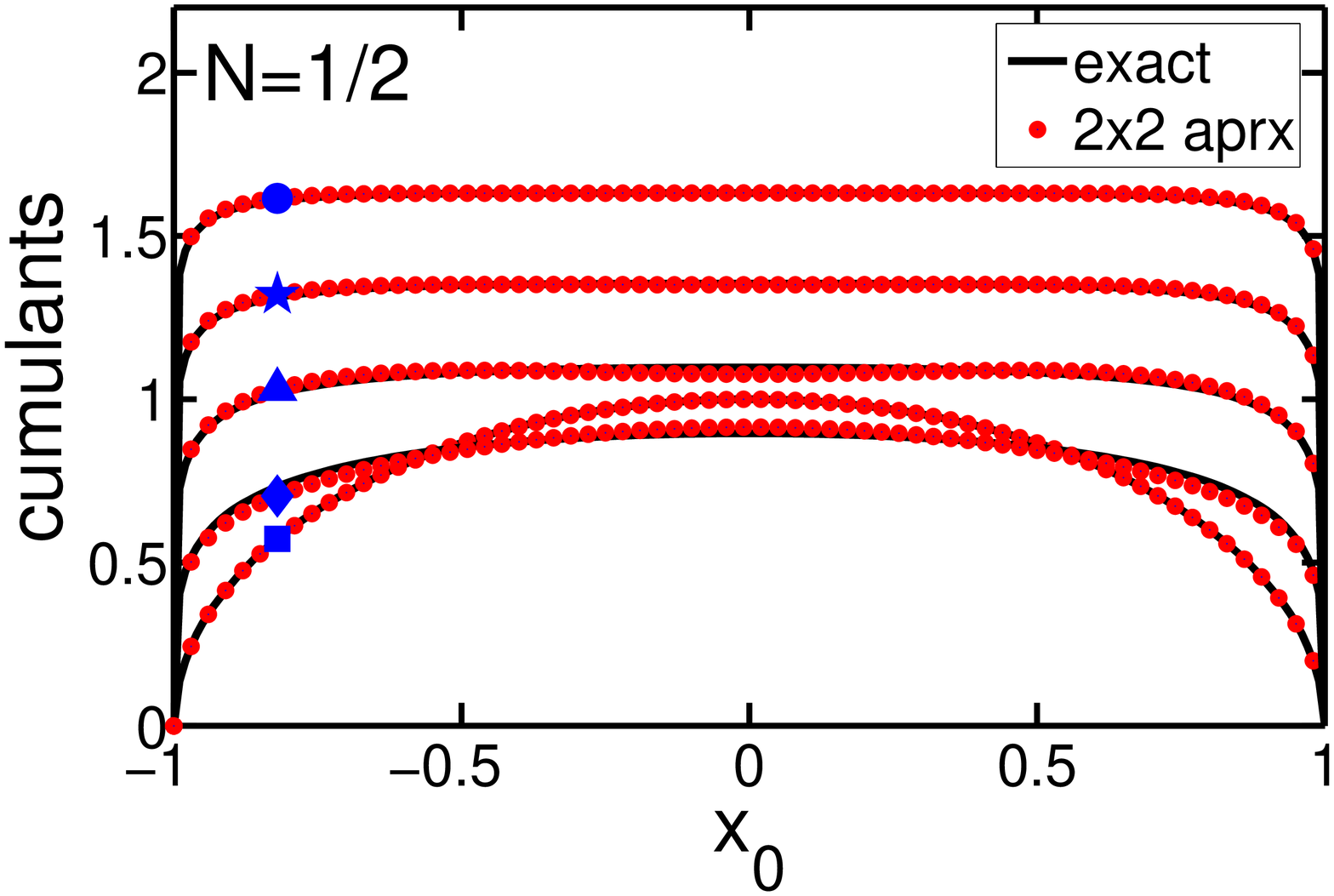}}
\centerline{\includegraphics[width=7cm]{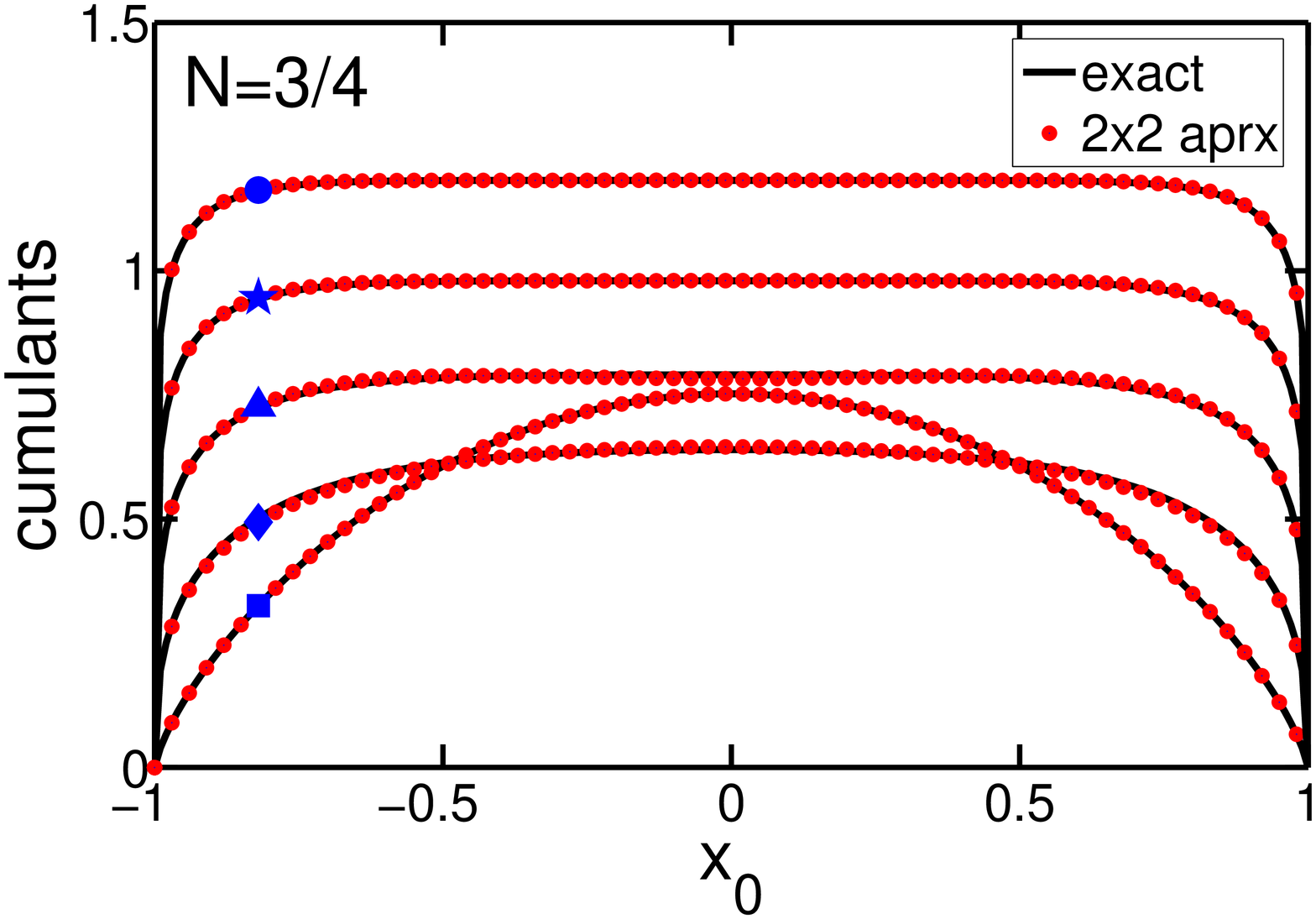}}
 \caption{A comparison of the first five cumulants as function of $x_0$ between
the exact numerical result and the $2 \times 2$ approach for
$N=1/4$, $N=1/2$ and $N=3/4$. The symbols $\blacksquare$,
$\blacklozenge$, {\large$\blacktriangle $}, $\bigstar$, and
{\Large$\bullet$} correspond to $\langle t \rangle$,
$\kappa^{1/2}_2$, $\kappa^{1/3}_3$, $\kappa^{1/4}_4$ and
$\kappa^{1/5}_5$ respectively.}
 \label{fig:cumulants}
\end{figure}

We calculated numerically the moments using both
Eq.~(\ref{eq:solmom}) and a direct numerical diagonalization of
$(-\Delta)^N$ as given by Eq.~(\ref{eq:matrix1}). In
Fig.~\ref{fig:cumulants}, we present the cumulants of the
distribution $\kappa_m(x_0)$ for $N=1/4,1/2,3/4$. Recall that the
cumulants $\kappa_m$ are simple combinations of $\langle t^m\rangle$
\cite{abram}, and that for $m=1,2$ they coincide with the mean and
the variance respectively. We also present results obtained using
only the first term in Eq.~(\ref{eq:solmom}), namely the ground
state, which we calculate here using the $2 \times 2$ approximation
(\ref{eq:anyNlam0},\ref{eq:anyNv0}). As can be seen in
Fig.~\ref{fig:cumulants} the agreement between the two approaches is
very good. This means that knowledge of the ground state provides a
reasonable description of the distribution. Note that the cumulants
grow faster as the value of $N$ decreases.

\begin{figure}[ht]
\centerline{\includegraphics[width=7cm]{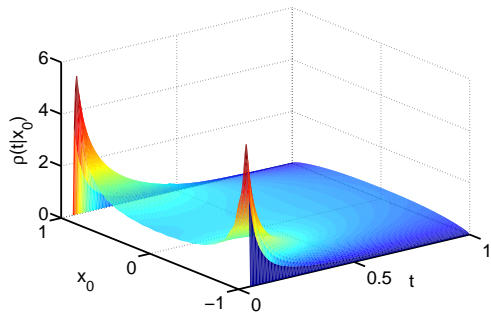}}
 \caption{The distribution of FPT $\rho(t|x_0)$ as a function of $x_0$ and $t$ for $N=1/2$.}
 \label{fig:PDF05}
\end{figure}

However, these results show that $\kappa_m^{1/m}$ increases with $m$
for all values of $x_0$, implying that the mean $\langle
t\rangle\equiv\kappa_1$ is generically not representative, since the
distribution is not peaked around it. In that case the variance, and
other higher order cumulants, are necessary to characterize the full
distribution. Put differently, statistical inference regarding first
passage properties based only on the mean can lead to wrong
conclusions since the errors bars, calculated using the higher order
moments, are typically larger than the the mean value itself. Last,
Fig.~\ref{fig:PDF05} shows the full distribution of FPT,
$\rho(t|x_0)$, for $N=1/2$, calculated using Eq.~(\ref{eq:solrho})
and a straightforward application of the numerical scheme presented
above. For long times $t\gg1$ the tail of $\rho(t|x_0)$ is always
exponential, and the lowest eigenvalue $\lambda_0(N)$ controls its
decay rate. However, for short times, many terms has to be retained
in the sum (\ref{eq:solrho}) giving rise to a nontrivial form.

\section{Summary and discussion}

In this letter we obtain both analytical and numerical information
about the spectrum of the fractional Laplacian in bounded domains,
extending previous results for integer powers of the Laplacian
\cite{Del^N}. These results allow us to address the timely question
of FPT statistics. We show that in general the mean first passage
time does not provide a proper description of the FPT distribution
since the variance (as well as higher order cumulants) become more
and more important as the L\'evy stability index $N$ becomes smaller
(see Fig.~\ref{fig:cumulants}). This means that the distribution is
not peaked around the average, as often implied when only the
average is discussed. Therefore, the MFPT is not representative of
the distribution, and the whole distribution $\rho(t|x_0)$ has to be
retained. A direct consequence of these results is that the
distribution of FPT is not Gaussian, contrarily to what is
implicitly assumed when focussing on the average (MFPT) only. We
show how to obtain the FPT distribution numerically for any $N$. In
addition we show that a simple ``$2\times 2$" approximation is able
to reproduce accurately the cumulants. Our results (\ref{eq:solrho})
suggest that the tail of the distribution is exponential, while for
short times it has a complex nontrivial form.

An interesting point, is that the lowest eigenvalue $\lambda_0$
becomes smaller than one (see the inset in Fig.~\ref{fig:lambdas})
in a whole range of small $N$'s. Even though this result does not
affect the FPT statistics studied here, this might have important
consequences in systems described by the Fractional Laplacian, whose
stability is determined by large powers of the eigenvalue of the
ground state (an example of such a system can be found
in~\cite{Cerda}). Our result suggests that in such cases tuning the
value of $N$ (by changing the fractality of the medium for example)
can control the stability of the system.

Note that we were interested above only in the even part of the
spectrum of $(-\Delta)^N$ (that is the even eigenfunctions). The
reason is that for the statistics of FPT with two absorbing
boundaries we need to know only the even spectrum since the odd
eigenfunction do not contribute to the expression for $\rho(t|x_0)$ as given by
Eq.~(\ref{eq:solrho}) due to the integral there. However, information
about the odd spectrum can be easily obtained by replacing $j
\rightarrow (j+1/2)$ in Eqs.~(\ref{eq:f}--\ref{eq:Dj}).

Last, we comment on how to treat other boundary conditions, such as
reflecting or mixed ones \cite{Dybiec}. The key observation is that
one has $C_j^{\left( {2N + \frac{1}{2}} \right)} (x) = P_j^{(2N,2N)}
(x)$ where $P_j^{\left( {\mu ,\nu } \right)} (x)$ is the $j^{th}$
Jacobi polynomial \cite{abram}. Using orthogonality properties and BC
of this polynomials it can be shown that the relevant basis for
mixed BC (i.e., one reflecting and one absorbing boundary) is proportional to
$\left\{ {\left( {1 + x} \right)^N \left( {1 - x} \right)^{N + 1}
P_j^{\left( {2N,2N + 2} \right)} \left( x \right)} \right\}$.

{\it Acknowledgements.}--This work was supported by EEC PatForm
Marie Curie action (E.K.). Laboratoire de Physique Statistique is associated with
Universities Paris VI and Paris VII.

\end{document}